\documentstyle[seceq,epsf]{ptptex}

\makeatletter

\def\lesssim{\mathrel{\mathpalette\vereq<}}
\def\vereq#1#2{\lower3pt\vbox{\baselineskip1.5pt \lineskip1.5pt
\ialign{$\m@th#1\hfill##\hfil$\crcr#2\crcr\sim\crcr}}}

\def\gtrsim{\mathrel{\mathpalette\vereq>}}

\def\alt{\lesssim}
\def\agt{\gtrsim}

\makeatother
\preprintnumber{OU-TAP-44\\ gr-qc/} 
\title{
Gravitational Waves by a Particle in Circular Orbits around a
 Schwarzschild black hole {{\em\it $-$  5.5 Post-Newtonian Formula $-$}} }
\author{Takahiro {\sc Tanaka}, 
Hideyuki {\sc Tagoshi}$^{*}$ and Misao {\sc Sasaki}}
\inst{
Department of Earth and Space Science, 
Osaka University, Toyonaka 560  \\
$^*$National Astronomical Observatory, 
Mitaka, Tokyo 181 }

\recdate{
}

\abst{Using the 
post-Newtonian (PN) expansion technique of the 
gravitational wave perturbation 
around a Schwarzschild black hole, we calculate 
analytically the energy flux of gravitational waves induced by 
a particle in circular orbits 
up to the 5.5PN order, i.e. $O(v^{11})$ beyond Newtonian. 
By comparing the formula with numerical data,
we find that the error of the 5.5PN formula  
is about $4\%$ when the particle is on the last stable circular 
 orbit. We also estimate the error $\Delta N$ in the total cycle of 
gravitational waves from coalescing compact binaries 
in a laser interferometer's band produced by using the post-Newtonian
approximations. We find that, 
as for the neutron star-black hole binaries, 
the 4.5PN approximation gives $\Delta N\alt1$ for
a black hole of mass $M<40M_\odot$, while it gives
$\Delta N\agt1$ for a black hole of mass $M>40M_{\odot}$. }

\begin{document}

\maketitle

\section{Introduction}
Gravitational waves from coalescing compact binaries are 
the most promising candidates which 
will be able to be detected by the near-future, 
ground based laser interferometric 
gravitational wave detectors such as LIGO, VIRGO, TAMA, GEO600 etc.
\cite{ref:thorne}

If a neutron star or a small black hole spirals into a 
black hole with mass \\
$<300M_\odot$, the inspiral wave form 
will be detected by the above detectors. 
When a signal of gravitational waves is detected, 
we will try to extract parameters for binaries, such as 
masses and spins etc., from inspiral wave forms 
using the matched filtering technique. \cite{ref:three}
In this method, parameters for binaries are determined 
by cross-correlating the noisy signal from detectors 
with theoretical templates. 
If the signal and the templates lose phase with each other by 
one cycle over $\sim 10^3 -10^4$ cycles 
as the waves sweep through the interferometer band, their 
cross correlation will be significantly reduced. 
This means that, in order to extract the information optimally, 
we need to make theoretical templates 
which are accurate to better than one cycle during an entire sweep 
through the interferometer's band. \cite{ref:three}

The standard method to calculate inspiraling wave forms 
from coalescing binaries is the post-Newtonian expansion of the Einstein 
equations, in which the orbital velocity $v$ of binaries 
is assumed to be small compared to the speed of light. 
Although the post-Newtonian calculation technique is being
developed to apply to higher order calculation, 
\cite{ref:postnewton} 
it becomes more and more difficult and complicated. 
Thus, it would be very helpful if we could have another 
reliable method to calculate the high order post-Newtonian 
corrections. 

As an alternative method, the post-Newtonian expansion 
of the black hole perturbation has been developed.
There, one considers gravitational waves from a particle 
of mass $\mu$ orbiting a black hole of mass $M$, assuming 
$\mu \ll M$. 
Although this method is restricted to the case $\mu\ll M$, 
we can calculate very high-order post-Newtonian corrections to 
gravitational waves by means of a relatively simple analysis, 
compared to the standard post-Newtonian analysis. 

Since LIGO and VIRGO will be able to detect 
gravitational waves from binaries of mass 
less than $\sim 300 M_{\odot}$, it is important to 
construct templates for 
such  binaries. The frequency of gravitational waves for such massive 
binaries, however, comes into the frequency band for LIGO and VIRGO at 
$r/M$ $\sim 16(100M_{\odot}/M)^{2/3}$, i.e., the highly relativistic region. 
We do not know whether 
the convergence property of the post-Newtonian approximation is good or not 
for such a highly relativistic orbit. 
As was reported by Tagoshi and Sasaki, \cite{ref:TS} the post-Newtonian 
convergence of the total orbital phase during the detectable frequency band 
for LIGO and VIRGO can be very slow for these binaries. 
Hence, it is an urgent problem to clarify 
the extent to which the convergence property of the post-Newtonian 
expansion is good. For this purpose, 
we study the energy loss rate of the binary up to $v^{11}$ order 
in this paper, where $v:=(M/r)^{1/2}$. 

This direction of research was first pursued done analytically by 
Poisson \cite{ref:poisson} 
to $O(v^3)$ and numerically by Cutler et al. \cite{ref:three}
to $O(v^5)$. 
A highly accurate numerical calculation was done by 
Tagoshi and Nakamura, \cite{ref:TN} and they gave a fitting formula 
to $O(v^8)$. They found the appearance of $\ln v$ terms 
in the energy flux at $O(v^6)$ and at $O(v^8)$. 
They also clarified that an accuracy of the energy flux to at least
$O(v^6)$ is needed to construct template wave forms for coalescing 
binary neutron stars. 
An analytical calculation to the same order was done by
Tagoshi and Sasaki \cite{ref:TS}, based on the 
formulation built up by Sasaki, \cite{ref:MS} and they 
confirmed the result of Tagoshi and Nakamura. 
Recently, a systematic method to calculate the higher order 
corrections was developed by Mano, Suzuki and Takasugi.\cite{ref:MST} 

Note that, both in these analyses and in this paper, 
the particle is treated as 
a test particle. According to the order of magnitude 
estimation, the quadrupole moment of a star produced by the 
tidal deformation of the star affects gravitational waves 
at $O(v^{10})$. \cite{ref:tidal} Thus we cannot regard the 
realistic compact star as a test particle. 
Nevertheless, we ignore this fact and treat
the star as an ideal structureless particle in this paper, 
expecting that we can still obtain useful information 
by analyzing such an ideal system.

This paper is organized as follows. 
In  \S 2, we give the general formulas and conventions used in 
this paper very briefly. The full description of the formalism can be
found in Tagoshi and Sasaki \cite{ref:TS} and references therein. 
In \S 3, we show the expressions of the energy flux 
extended to $O(v^{11})$. 
In \S 4, we discuss the accuracy of the post-Newtonian
expansion. 
\S 5 is devoted to a summary. 

Throughout this paper we use the units such that $c=G=1$. 

\section{General formulation}

 For the reader's convenience, we attach a minimum amount of the explanation 
about the black hole perturbation formalism in order to make the convention 
definite. 
For the reader who would like to know the details, see the 
reference \cite{ref:TS} (Some typographical errors in the formulas 
in this reference are corrected here). 
We consider the case in which a particle of small mass $\mu$ travels a circular 
orbit around a Schwarzschild black hole of mass $M\gg \mu$. 

 To calculate the gravitational luminosity, 
we consider the inhomogeneous Teukolsky equation, 
\begin{equation}
 \left[\Delta^2{d\over dr}\left({1\over \Delta}{d\over dr}\right)+U(r)\right]
  R_{\ell m\omega}(r) = T_{\ell m\omega}(r),
\label{Teu}
\end{equation}
where
\begin{equation}
 U(r)={r^2\over\Delta}\left[\omega^2 r^2
-4i\omega(r-3M)\right]-(\ell-1)(\ell +2), 
\quad \Delta=r(r-2M),
\end{equation}
and $T_{\ell m\omega}$ is the source term which reflects 
the energy momentum tensor of the small particle. 
We omit the explicit form of $T_{\ell m\omega}(r)$ here. 

We solve Eq. (\ref{Teu}) using the Green function method. For this purpose, we 
need a homogeneous solution $R_{\ell\omega}^{in}$ of Eq. (\ref{Teu}) which 
satisfies the boundary conditions 
\begin{equation}
R_{\ell\omega}^{in}=\left\{
  \begin{array}{lcc}
    D_{\ell\omega}\Delta^2 e^{-i\omega r^{*}} & \hbox{for} 
    & r^{*}\rightarrow -\infty, \\
    r^3 B_{\ell\omega}^{out} e^{i\omega r^{*}} +
    r^{-1} B_{\ell\omega}^{in}  e^{-i\omega r^{*}} & \hbox{for} 
    & r^{*}\rightarrow +\infty,
   \end{array}
\right.
\end{equation}
where $r^{*}=r+2M\ln(r/2M -1)$. 
Then the outgoing-wave solution of Eq. (\ref{Teu})
at infinity with appropriate boundary conditions at horizon is given by 
\begin{eqnarray}
 R_{\ell m\omega}(r\rightarrow\infty)  = & & {r^3 e^{i\omega r^{*}}\over 
    2i\omega B_{\ell \omega}^{in}}\int_{2M}^{\infty} dr R_{\ell \omega}^{in}
    T_{\ell m\omega}(r)\Delta^{-2}
\cr  
   =:&& r^3 e^{i\omega r^{*}}\tilde Z_{\ell m\omega}.
\label{eq4}
\end{eqnarray}
In the case of a circular orbit, the specific energy $\tilde E$ and 
angular momentum $\tilde L$ of the particle are given by 
\begin{equation}
\tilde E=(r_{0}-2M)/\sqrt{r_0(r_0-3M)},
\label{eq5}
\end{equation}
and
\begin{equation}
\tilde L=\sqrt{M r_{0}}/\sqrt{1-3M/r_0},
\label{eq6}
\end{equation}
where $r_0$ is the orbital radius. The angular frequency is given by 
$\Omega=(M/r_0^3)^{1/2}$. 
Defining $_s b_{\ell m}$ by 
\begin{eqnarray}
_0 b_{\ell m} & = & {1\over2}\left[(\ell-1)\ell(\ell+1)(\ell+2)\right]^{1/2}
    {}_{0} Y_{\ell m}\left({\pi\over 2},0\right)\tilde E r_0/(r_0-2M),
\cr
_{-1} b_{\ell m} & = & \left[(\ell-1)(\ell+2)\right]^{1/2}
    {}_{-1} Y_{\ell m}\left({\pi\over 2},0\right)\tilde L/r_0,
\cr 
_{-2} b_{\ell m} & = & _{-2} Y_{\ell m}\left({\pi\over 2},0\right)
            \tilde L\Omega,
\label{eq8}
\end{eqnarray}
where $_{n} Y_{\ell m}(\theta,\varphi)$ are the spin-weighted spherical 
harmonics, 
$\tilde Z_{\ell m\omega}$ is found to take the form 
\begin{equation}
 \tilde Z_{\ell m\omega}=Z_{\ell m}\delta(\omega-m\Omega),
\label{eq9}
\end{equation}
where
\begin{eqnarray}
Z_{\ell m}& = &{\pi\over i\omega r_0^2 B_{\ell\omega}^{in}}
  \Biggl\{\Biggl[
      -_0 b_{\ell m} -2i _{-1}b_{\ell m}
             \left(1+{i\over 2}\omega r_0^2/(r_0-2M)\right)
\cr &&
    + i _{-2} b_{\ell m}\omega r_0 (1-2M/r_0)^{-2}
     \left(1-M/r_0+{1\over 2}i\omega r_0\right)\Biggr] R_{\ell m}^{in}
\cr &&
    +\left[i_{-1}b_{\ell m}-_{-2}b_{\ell m}
         \left(1+i\omega r_0^2/(r_0-2M)\right)\right] 
           r_0 {R^{in}_{\ell\omega}}'(r_0)
\cr &&
      +{1\over 2} {}_{-2} b_{\ell m} r_0^2 {R_{\ell \omega}^{in}}''(r_0)
      \Biggr\}\quad.
\label{eq10}
\end{eqnarray}
In terms of the amplitudes $Z_{\ell m}$, the gravitational wave 
luminosity is given by
\begin{equation}
 {dE\over dt}=\sum_{\ell=2}^{\infty} \sum_{m=-\ell}^{\ell}
   \vert Z_{\ell m}\vert^2/2\pi \omega^2,
\end{equation}
where $\omega =m\Omega$. 
We calculate this quantity in the post-Newtonian expansion, that is, 
in the expansion with respect to $v=(M/r)^{1/2}$. 
Thus the only remaining task is to calculate 
the series expansion of the ingoing-wave Teukolsky function 
$R_{\ell \omega}^{in}$ in terms of $\epsilon:=2M\omega=O(v^3)$ 
and $z:=\omega r=O(v)$ and the Wronskian $B^{in}_{\ell m}$ 
in terms of $\epsilon$.

\section{The gravitational wave luminosity and the 
accuracy of the post-Newtonian expansion} 

As noted in the introduction, a new method to calculate the 
Teukolsky function in the post-Newtonian expansion was constructed by 
Mano, Suzuki and Takasugi. \cite{ref:MST} 
However, we have derived the following results by using the 
formalism developed by Sasaki, \cite{ref:MS} and 
the details of derivation is too complicated and 
seem not worths of note here, because a new more powerful 
method has been developed. Thus we present only the final results 
of the gravitational wave luminosity. 
The formulas for $R_{\ell \omega}^{in}$ and  $B^{in}_{\ell m}$ 
and the luminosity of each mode are given in Appendix A. 

The total luminosity becomes as \cite{ref:tago}
{\jot0pt
\begin{eqnarray}
{dE\over dt}=&&\left({dE\over dt}\right)_N \Biggl[
1 - {{1247}\over {336}}v^2 + 4\,\pi \,{v^3} 
  - {{44711}\over {9072}}v^4 
-{{8191\,\pi }\over {672}} v^5 
\cr &&\cr &&\vspace{6pt}
+  \left( {{6643739519}\over {69854400}} - {{1712\,{\gamma}}\over {105}} 
- {{1712\,{\ln v}}\over {105}} + {{16\,{{\pi }^2}}\over 3} - 
     {{3424\,\ln 2}\over {105}} \right) v^6  
- {{16285\,\pi }\over {504}} v^7
\cr &&\cr &&\vspace{6pt}
+  \Biggl( -{{323105549467}\over {3178375200}} + 
     {{232597\,{\gamma}}\over {4410}} + {{232597\,{\ln v}}\over {4410}} 
\cr  && \hspace{4cm}
   - {{1369\,{{\pi }^2}}\over {126}} + {{39931\,\ln 2}\over {294}} - 
     {{47385\,\ln 3}\over {1568}} \Biggr) v^8 
\cr  &&\cr &&\vspace{6pt}
+  \left( {{265978667519\,\pi }\over {745113600}} - 
     {{6848\,{\gamma}\,\pi }\over {105}} - 
    {{6848\,{\ln v}\,\pi }\over {105}} 
  - {{13696\,\pi \,\ln 2}\over {105}} \right) v^9 
\cr &&\cr &&\vspace{6pt}
 + \Biggl( -{{2500861660823683}\over {2831932303200}} 
  + {{916628467\,{\gamma}}\over {7858620}} 
  + {{916628467\,{\ln v}}\over {7858620}} 
  - {{424223\,{{\pi }^2}}\over {6804}} 
\cr  && \hspace{4cm}
  - {{83217611\,\ln 2}\over {1122660}} 
  + {{47385\,\ln 3}\over {196}}  \Biggr) v^{10}
\cr  &&\cr &&\vspace{6pt}
  +\Biggl( {{8399309750401\,\pi }\over {101708006400}} 
  + {{177293\,{\gamma}\,\pi }\over {1176}} 
  + {{177293\,{\ln v}\,\pi }\over {1176}} 
\cr  && \hspace{5cm}
  + {{8521283\,\pi \,\ln 2}\over {17640}} 
  - {{142155\,\pi \,\ln 3}\over {784}}\Biggr) v^{11} \Biggr]
\cr &&\cr
=&&\left({dE\over dt}\right)_N \Biggl[
 1 - 3.711309523809524\,{v^2} + 12.56637061435917\,{v^3} 
\cr &&\hspace{1.5cm}
- 4.928461199294533\,{v^4} 
  - 38.29283545469344\,{v^5} 
\cr &&\cr &&\hspace{1.5cm}\vspace{2pt}
+ \left( 115.7317166756113 - 
   16.3047619047619\,{\ln v}
      \right) \,{v^6} 
\cr &&\cr &&\hspace{1.5cm}\vspace{6pt}
 - 101.5095959597416\,{v^7} + 
\cr &&\cr &&\hspace{1.5cm}\vspace{6pt}
  \left( -117.5043907226773 + 52.74308390022676\,{\ln v} \right) \,{v^8}  
\cr &&\cr &&\hspace{1.5cm}\vspace{6pt} 
  +  \left( 719.1283422334299 - 204.8916808741229\,{\ln v} \right) 
\,{v^9}  
\cr &&\cr && \hspace{1.5cm}\vspace{6pt} 
  +  \left( -1216.906991317042 + 116.6398765941094\,{\ln v} \right) 
\,{v^{10}} 
\cr && \hspace{1.5cm} 
+  \left( 958.934970119567 + 473.6244781742307\,{\ln v} \right) 
\,{v^{11}}+ 
\cdots \Biggr]\quad. 
\label{formula} 
\end{eqnarray}} 

Using the above results, 
we compare the formula for the gravitational 
wave flux with numerical results and investigate 
the accuracy of the post-Newtonian expansion. 

A high precision numerical calculation 
of gravitational waves from a particle 
in a circular orbit around a Schwarzschild black hole 
has been performed by Tagoshi and Nakamura. \cite{ref:TN} 
Since no assumption was made about the velocity of the test particle, 
their results are correct relativistically in the limit 
$\mu\ll M$. 
In that work, $dE/dt$ was calculated only for $\ell=2\sim 6$. 
Then, for the orbital radius $r_0\le 100 M$, 
we calculate $dE/dt$ again for all modes of $\ell=2\sim 6$
and for $\ell=7$ with odd $m$. 
The estimated accuracy of the calculation is about $10^{-11}$, which 
turns out to be accurate enough for the present purpose. 
As for the radius $r_0>100 M$, since it is expensive to calculate $dE/dt$, 
we use the data calculated by Tagoshi and Nakamura \cite{ref:TN}
which contain modes from $\ell=2$ to $6$. 

\begin{figure}[htb]
\vspace{4cm}
{\footnotesize 
Figs. 1 and 2 ~The error of the post-Newtonian formulas as a
function  of the Schwarzschild coordinate radius $r$ for 
$6\leq r/M$ $\leq 100$ (Fig. 1) and $100\leq r/M$ $< 5000$
(Fig. 2). In Fig.~1, contributions from $\ell=2$ to $7$
are included. In Fig.~2, contributions only from $\ell=2$ to $6$ 
are included in both the post-Newtonian formulas and in the numerical 
data. }
\end{figure}

In Figs. 1 and 2, we show the error in the post-Newtonian formulas as 
a function of the orbital radius $r$.  
We refer to the post-Newtonian formula which include terms up to the order 
$v^n$ as the $(n/2)$PN formula. 
The error of the post-Newtonian formula is defined as 
\begin{equation}
{\rm error}=\left|1-\left(dE\over dt\right)_{n}\bigg/
\left(dE\over dt\right)\right|, 
\end{equation}
where $(dE/dt)_{n}$ and $(dE/dt)$ denote the $(n/2)$PN
formula and the numerical result, respectively. 
In the plot of Fig.~2, only the contributions from $\ell=2$ to $6$ 
are included in both the post-Newtonian formulas 
and the numerical data. 
We can see that, at small radius less than $r\sim 10M$, the error of 
the 1PN and 2.5PN formulas are larger than the 
other formulas. 
On the other hand, the Newtonian and the 2PN formulas
are very accurate within this radius. This is because those formulas 
coincide with the exact one accidentally at a radius between $6M$ and $10M$. 
The error of each post-Newtonian formula at the inner most stable circular 
orbit, $r=6 M$, becomes as follows;
$12\%$ (Newtonian), $66\%$ (1PN), $8.6\%$ (1.5PN), $3.4\%$ (2PN), 
$42\%$ (2.5PN), $11\%$ (3PN),
$5.4\%$ (3.5PN), $17\%$ (4PN), $8.4\%$ (4.5PN), 
$6.5\%$ (5PN), $4.1\%$ (5.5PN). 
As is expected, 
the errors of the post-Newtonian formulas 
decrease almost linearly up to $r\sim 5000 M$ in a log-log plot. 
This fact also suggests that the 
numerical data have accuracy of at least $\sim 10^{-18}$ at $r\sim 5000M$. 

In order to examine exactly to what order the post-Newtonian 
formulas are needed to do accurate estimation of 
the parameters of a binary using data from laser interferometers, 
we must evaluate the systematic error produced by incorrect templates, 
\cite{ref:CF}
However, here we simply 
calculate the total cycle of gravitational waves from a coalescing 
binary in a laser interferometer band and evaluate the error 
produced by the post-Newtonian formulas.
It has been suggested that 
whether the error in the total cycle is less than unity or not 
gives a 
useful guideline to examine the accuracy of the post-Newtonian 
formulas as templates \cite{ref:three} 
(see also \cite{ref:poisson2}). 

We ignore the finite mass effect in the post-Newtonian formulas and
interpret $M$ as the total mass and $\mu$ as the reduced mass 
of the system.
Further, since the error of the 5.5PN formula for $dE/dt$ is only a few 
percent even at $r=6M$, and since the error decreases monotonically as 
the orbital
radius increases, we regard the 5.5PN formula as if it were
the exact formula.

The total cycle $N$ of gravitational waves from an inspiraling 
binary is given by 
\begin{equation} 
N=\int^{v_i}_{v_f}dv {\Omega\over \pi} {dE/dv\over |dE/dt|}, 
\end{equation} 
where $v_i=(M/r_i)^{1/2}$, $v_f=(M/r_f)^{1/2}$, 
and $r_i$ and $r_f$ are the initial and final orbital separation of the 
binary. 
In the test particle limit, which we assume, $dE/dv$ is given by
\begin{equation}
{dE\over dv}=\mu {d\over dv}\left({{1-2v^2}\over(1-3v^2)^{1/2}}\right).
\end{equation} 
As noted before, we regard the 5.5PN formula (\ref{formula}) for $dE/dt$
as if it were the exact formula. 
We then introduce $P(v)$ $=(dE/dt)/(dE/dt)_N$ and $Q(v)$ 
$=(dE/dv)/(dE/dv)_N$, 
where $(dE/dv)_N=-\mu v$, and define the approximated total cycle
$N^{(n)}$ as 
\begin{equation} 
N^{(n)}=\int^{v_i}_{v_f}dv {\Omega\over \pi} {(dE/dv)_N\over (dE/dt)_N}
{Q_n(v)\over P_n(v)}, 
\end{equation} 
where $Q_n(v)$ and $P_n(v)$ are given by expanding $Q(v)$ and $P(v)$ 
in terms of $v$ and truncating them at order $v^n$. 
Note that we do not expand $Q_n(v)/P_n(v)$ in terms of $v$ further 
in calculating the above integral because such an expansion will produce 
additional errors in $N^{(n)}$. 
The error in the total cycle by using the
$(n/2)$PN formula is defined as 
\begin{equation} 
\Delta N^{(n)}=N^{(n)}-N.
\end{equation} 
Note that, because we have regarded the 5.5PN formula for $dE/dt$ to be
exact, $\Delta N^{(11)}$ is produced only by the error in $dE/dv$.

The results are listed in Table I. We find that
the post-Newtonian approximations
 with $n\le 5$ produce errors much greater than 1. 
This slow convergence was observed in previous works
\cite{ref:cutler,ref:TN,ref:TS,ref:SSTT,ref:TSTS}.
For a system of a neutron star with mass $1.4M\odot$ and 
a black hole with mass much larger than $1.4M_\odot$, 
the convergence becomes much slower.
Accidentally, the 3PN formula 
seems to be good irrespective of the masses of the binary. 
At $n=7$ and $8$,
the post-Newtonian approximations produce $\Delta N\sim 1$
for a binary with mass of each star less than about $10M_\odot$
but $\Delta N\sim10$ for the case where 
one of binary stars is a black hole with 
mass $\agt40M_\odot$.
At $n\ge 9$, $\Delta N\alt1$ for all the cases investigated here.

{}From these results, we can say that at least 
the 3PN formula is needed for template wave forms. 
The 4.5PN or higher order approximation seems to be
sufficient for constructing template wave forms.
It should be noted, however, if the black hole mass is
greater than $70M_\odot$, $\Delta N$ becomes greater than 2 
even in the case of the 5.5PN formula. 
Nevertheless, this does not necessarily mean that
the post-Newtonian approximations, even if we include the 5.5PN terms,
is inadequate for calculating the theoretical wave form for such a
system. To obtain a definite conclusion, one must perform a more
detailed analysis of the systematic error as done by Cutler and Flanagan. 
\cite{ref:CF}

\begin{table}
\caption{
The error $\Delta N^{(n)}$ in the total cycle
of gravitational waves from coalescing compact binaries 
in a laser interferometer's band produced by using the 
$(n/2)$PN formula. The initial frequency is 10Hz, 
which corresponds to the initial radius 
$r/M\simeq$ $347(M_\odot/M)^{2/3}$, 
and the final frequency is the one at which the binary is $r=6 M$.
Note that $\Delta N^{(11)}$ is produced only by the error in the 
formula for $dE/dv$. }
\begin{tabular}{|c|c|c|c|c|c|} 
\hline
$n$ &{\small (1.4$M_\odot$,1.4$M_\odot$)}&{\small (10$M_\odot$,10$M_\odot$)}
&{\small (1.4$M_\odot$,10$M_\odot$)}&{\small (1.4$M_\odot$,40$M_\odot$)}& 
{\small (1.4$M_\odot$,70$M_\odot$)}\\ \hline
0 & $-118$ &12       &9.6      &100  &140 \\
2 &  239   &67       &223      &312  &353 \\
3 &  10    &6.3      &17       &40   &56  \\
4 & $-1.3$ &1.2      &1.6      &13   &24  \\
5 &  11    &8.8      &22       &56   &78  \\
6 &$-0.17$ &$-0.018$ &$-0.19$  &0.75 &2.7 \\
7 &  1.1   &1.1      &2.5      &7.9  &13  \\
8 &  0.94  &0.91     &2.1      &6.7  &11  \\
9 &  0.11  &0.12     &0.25     &0.93 &1.7 \\
10& 0.20   &0.20     &0.46     &1.5  &2.6 \\
11& 0.17   &0.17     &0.39     &1.3  &2.2 \\  
\hline
\end{tabular}
\end{table}
One reason for the relatively slow convergence of 
the post-Newtonian expansion in calculating the cycle $N$ 
is that we are integrating to a small radius $r=6M$, which implies a 
value of $v\sim 1$. 
Further, we can understand some of the above results 
by a simple order of magnitude 
estimation (see paper TSTS, section IV). 
Since the cycles 
are mainly accumulated around $\sim 10$Hz, which is the 
lowest detectable frequency region of the laser interferometers such as 
LIGO and VIRGO, $N$ is 
approximately given by 
\begin{eqnarray} 
N \sim 1.9 \times 10^3 \left({10M_{\odot} \over M} \right)^{5/3}
\left({M \over 4\mu} \right), 
\end{eqnarray} 
where $M$ and $\mu$ are the total mass and reduced mass, respectively. 
This implies that the template must have an accuracy of at least
\begin{eqnarray}
\sim 5\times10^{-4} \left({M \over 10M_{\odot}} \right)^{5/3}
\left({4\mu \over M} \right), 
\end{eqnarray} 
when the frequency of gravitational wave becomes 10Hz. 
Then, for fixed $M$, 
we need much better accuracy in the case $\mu\ll M$ than in the case 
of equal mass, since $N$ becomes large. 
On the other hand, for fixed $\mu$, the required accuracy seems 
to become lower as $M$ becomes larger. 
However, since the non-dimensional radius $r/M$, 
at which  the frequency of gravitational waves becomes 10Hz, 
becomes smaller for larger $M$, the importance of higher order 
corrections increases. 
This is the reason why the convergence of the case 
(1.4$M_\odot$,70$M_\odot$) is slower than that of 
(1.4$M_\odot$,40$M_\odot$). 

\section{Summary} 
We have calculated the energy flux of gravitational waves 
induced by a particle of small mass in 
circular orbits around a Schwarzschild black hole up to $O(v^{11})$ 
beyond the quadrupole formula analytically. 
By comparing the analytically derived post-Newtonian formulas with 
numerical results, we discussed the accuracy of the post-Newtonian 
expansion.  We have found that the accuracy of the 5.5PN formula is 
good because 
the error of the $dE/dt$ formula is about $4\%$ even at the inner-most 
stable circular orbit $r=6M$. 

We have also calculated the total cycle of 
gravitational waves in a laser interferometer's band and estimated
the error produced by the post-Newtonian expansion.
We have found that in the case when each star of a 
binary has the same mass, the 4.5PN approximation gives a
sufficiently accurate cycle 
irrespective of the value of total mass.
On the other hand, in the case of 
neutron star-black hole binaries for which 
the mass of the black hole greater
than $\sim40M_\odot$, the convergence of the post-Newtonian 
expansion is slow. 
When the total mass is about 40$M_\odot$, 
the error is about 1 for the 5.5PN approximation.  
When the total mass is greater than $\sim 70M_\odot$, 
the error becomes greater than 2 even if we go to the 5.5PN order. 
It is difficult to conclude, however, whether the post-Newtonian
expansion can or cannot be used 
to construct theoretical templates for such systems 
only from the above results, 
because the value of $\Delta N$ is marginal.
We need a detailed analysis of the systematic error, in the 
parameter estimation, produced by 
the post-Newtonian expansion, such as the one done in Ref. 12). 
This is left for future work.

\begin{center}
\Large{\bf Acknowledgements}
\end{center}
We thank M. Shibata for discussions and for providing us his 
numerical data of gravitational waves. 
H.T. was supported by 
Japan Society for the Promotion of Science.
He was also supported in part 
by NSF Grant No. AST-9417371 and NASA Grand No. NAGW-4268 of U.S.A. 
when he stayed at Caltech where some parts of this work were done. 
This work was also supported in part by the Japanese Grant-in-Aid for 
Scientific Research of the Ministry of Education, 
Science, Sports, and Culture, No. 04234104. 

\appendix
\section{Formulas for $R_{\ell \omega}^{in}$ and  $B^{in}_{\ell m}$ and 
the gravitational wave luminosity}

The post-Newtonian expansion of the ingoing 
Teukolsky function in the near zone, 
where $z$ is small, is given by 
\begin{eqnarray}
R^{in}_{2 \omega}=
&& 
\Biggl({{4\,{z^4}}\over 5} + {{8\,i}\over {15}}\,{z^5} 
  - {{22\,{z^6}}\over {105}} - 
    {{2\,i}\over {35}}\,{z^7} + {{23\,{z^8}}\over {1890}} + 
    {{2\,i}\over {945}}\,{z^9} - {{13\,{z^{10}}}\over {41580}} 
\cr && \quad - 
    {i\over {24948}}\,{z^{11}} + {{59\,{z^{12}}}\over {12972960}} + 
    {i\over {2162160}}\,{z^{13}} - {{83\,{z^{14}}}\over {1945944000}} - 
    {i\over {277992000}}\,{z^{15}}\Biggr) 
\cr &&+ 
  \Biggl( {{-8\,{z^3}}\over 5} - {{3\,i}\over 5}\,{z^4} - 
     {{8\,{z^5}}\over {63}} - {{13\,i}\over {90}}\,{z^6} + 
     {{109\,{z^7}}\over {1890}} + {{341\,i}\over {22680}}\,{z^8} 
\cr && \quad - 
     {{9403\,{z^9}}\over {3118500}} - {{293\,i}\over {594000}}\,{z^{10}} + 
     {{38963\,{z^{11}}}\over {567567000}} + 
     {{75529\,i}\over {9081072000}}\,{z^{12}} \Biggr) \,\epsilon  
\cr && + 
  \Biggl( {{4\,{z^2}}\over 5} + {{123317\,{z^4}}\over {36750}} + 
     {{231479\,i}\over {110250}}\,{z^5} - {{889954\,{z^6}}\over {1157625}} - 
     {{454499\,i}\over {2315250}}\,{z^7} 
\cr && \quad+ 
     {{215321483\,{z^8}}\over {5501034000}} + 
     {{35106811\,i}\over {5501034000}}\,{z^9} - 
     {{214\,{z^4}\,\ln z}\over {525}} 
- {{428\,i}\over {1575}}\,{z^5}\,\ln z 
\cr && \quad 
+ {{1177\,{z^6}\,\ln z}\over {11025}} + 
     {{107\,i}\over {3675}}\,{z^7}\,\ln z - 
     {{2461\,{z^8}\,\ln z}\over {396900}} - 
     {{107\,i}\over {99225}}\,{z^9}\,\ln z \Biggr) \,{{\epsilon }^2} 
\cr && + 
  \Biggl( {{-66823\,{z^3}}\over {12250}} - {{99851\,i}\over {55125}}\,{z^4} - 
     {{504569\,{z^5}}\over {694575}} - {{2488639\,i}\over {3969000}}\,{z^6} + 
     {{428\,{z^3}\,\ln z}\over {525}} 
\cr && \quad + 
     {{107\,i}\over {350}}\,{z^4}\,\ln z + 
     {{428\,{z^5}\,\ln z}\over {6615}} 
   + {{1391\,i}\over {18900}}\,{z^6}\,\ln z \Biggr) \,{{\epsilon }^3} 
\cr && \quad + 
  \left( {{471487\,{z^2}}\over {220500}} - {{263\,i}\over {1260}}\,{z^3} - 
     {{214\,{z^2}\,\ln z}\over {525}} \right) \,{{\epsilon }^4} ,
\cr
R^{in}_{3\omega}=
&&
\Biggl(
 {{4\,{z^5}}\over {21}} + {{2\,i}\over {21}}\,{z^6} - {{2\,{z^7}}\over {63}} - 
    {i\over {135}}\,{z^8} + {{29\,{z^9}}\over {20790}} + 
    {i\over {4620}}\,{z^{10}} - {{47\,{z^{11}}}\over {1621620}} 
\cr && \quad\quad - 
    {i\over {294840}}\,{z^{12}} + {{23\,{z^{13}}}\over {64864800}} + 
    {i\over {29937600}}\,{z^{14}}\Biggr) 
\cr && + 
  \Biggl( {{-10\,{z^4}}\over {21}} - {{53\,i}\over {315}}\,{z^5} + 
     {{{z^6}}\over {210}} - {i\over {90}}\,{z^7} + 
     {{751\,{z^8}}\over {155925}} + {{1483\,i}\over {1247400}}\,{z^9} 
\cr &&\quad- 
     {{23\,{z^{10}}}\over {102960}} - {{367\,i}\over {10810800}}\,{z^{11}}
      \Biggr) \,\epsilon  
\cr && + \Biggl( {{8\,{z^3}}\over {21}} + 
     {i\over {14}}\,{z^4} + {{40337\,{z^5}}\over {46305}} + 
     {{79099\,i}\over {185220}}\,{z^6} - {{12562147\,{z^7}}\over {91683900}} - 
     {{4840537\,i}\over {157172400}}\,{z^8} - 
\cr &&\quad
     {{26\,{z^5}\,\ln z}\over {441}} - 
     {{13\,i}\over {441}}\,{z^6}\,\ln z + 
     {{13\,{z^7}\,\ln z}\over {1323}} + 
     {{13\,i}\over {5670}}\,{z^8}\,\ln z \Biggr) \,{{\epsilon }^2} 
\cr && + 
  \left( {{-2\,{z^2}}\over {21}} - {{182981\,{z^4}}\over {92610}} - 
     {{3753697\,i}\over {5556600}}\,{z^5} + 
     {{65\,{z^4}\,\ln z}\over {441}} + 
     {{689\,i}\over {13230}}\,{z^5}\,\ln z \right) \,{{\epsilon }^3}, 
\cr
R^{in}_{4\omega}=
&&
\Biggl({{2\,{z^6}}\over {63}} + {{4\,i}\over {315}}\,{z^7} - 
    {{13\,{z^8}}\over {3465}} - {{8\,i}\over {10395}}\,{z^9} 
\cr &&\quad+ 
    {{71\,{z^{10}}}\over {540540}} + {i\over {54054}}\,{z^{11}} - 
    {{37\,{z^{12}}}\over {16216200}} - {i\over {4054050}}\,{z^{13}}\Biggr) 
\cr &&+ 
  \Biggl( {{-2\,{z^5}}\over {21}} - {{4\,i}\over {135}}\,{z^6} + 
     {{142\,{z^7}}\over {51975}} - {{31\,i}\over {51975}}\,{z^8} + 
     {{929\,{z^9}}\over {2702700}} + {{8\,i}\over {96525}}\,{z^{10}} \Biggr) \,
   \epsilon  
\cr &&
   + \Biggl( {{5\,{z^4}}\over {49}} + {{97\,i}\over {4410}}\,{z^5} + 
     {{958223891\,{z^6}}\over {6051137400}} + 
     {{239560304\,i}\over {3781960875}}\,{z^7} 
\cr &&\quad
 - {{1571\,{z^6}\,\ln z}\over {218295}} - 
     {{3142\,i}\over {1091475}}\,{z^7}\,\ln z \Biggr) \,{{\epsilon }^2} 
\cr && + 
  \Biggl( {{-20\,{z^3}}\over {441}} - {i\over {196}}\,{z^4} \Biggr) \,
   {{\epsilon }^3} ,
\cr
R^{in}_{5\omega}=
&&\Biggl({{2\,{z^7}}\over {495}} + {{2\,i}\over {1485}}\,{z^8} - 
    {{7\,{z^9}}\over {19305}} - {i\over {15015}}\,{z^{10}} + 
    {{17\,{z^{11}}}\over {1621620}} + {i\over {737100}}\,{z^{12}}\Biggr) 
\cr &&+ 
  \Biggl( {{-7\,{z^6}}\over {495}} - {{67\,i}\over {17325}}\,{z^7} + 
     {{1831\,{z^8}}\over {4054050}} - {{43\,i}\over {4054050}}\,{z^9} \Biggr) 
    \,\epsilon  
\cr &&
   + \Biggl( {{28\,{z^5}}\over {1485}} + 
     {{59\,i}\over {14850}}\,{z^6} \Biggr) \,{{\epsilon }^2} ,
\cr
R^{in}_{6\omega}=
&&\Biggl({{8\,{z^8}}\over {19305}} + {{16\,i}\over {135135}}\,{z^9} - 
    {{4\,{z^{10}}}\over {135135}} - {{2\,i}\over {405405}}\,{z^{11}}\Biggr) 
\cr &&+ 
  \Biggl( {{-32\,{z^7}}\over {19305}} - {{163\,i}\over {405405}}\,{z^8} \Biggr)
    \,\epsilon  ,
\cr
R^{in}_{7\omega}=
&& 
{{8\,{z^9}}\over {225225}} + {{2\,i}\over {225225}}\,{z^{10}},
\end{eqnarray}
and the Wronskian for $\ell=2,$ and $3$ is given by 
\begin{eqnarray}
B^{in}_{2\omega}&=&{i\over \omega^2}e^{-i\epsilon(\ln 2\epsilon+\gamma)}
     \left(3-{3\over 4} i\epsilon\right)
   \Biggl[ 1 + \epsilon \,\left( {{5\,i}\over 3} - {{\pi }\over 2} - 
     i\,\left( \gamma  + \ln 2 \right)  \right)  
\cr 
&+& {{\epsilon }^2}\,\left( {{-457\,i}\over {420}}\,\pi  + 
     {{5\,{{\pi }^2}}\over {24}} + 
     {{457\,\left( \gamma  + \ln 2 \right) }\over {210}} + 
     {i\over 2}\,\pi \,\left( \gamma  + \ln 2 \right)  - 
     {{{{\left( \gamma  + \ln 2 \right) }^2}}\over 2} \right)  
\cr 
&+& 
{{\epsilon }^3}\,\Biggl( {{343\,i}\over {216}} + {{107\,\pi }\over {252}} + 
     {{491\,i}\over {1260}}\,{{\pi }^2} - {{{{\pi }^3}}\over {16}} + 
     {i\over 3}\,{\zeta (3)} + {{107\,i}\over {126}}\,
      \left( \gamma  + \ln 2 \right)  
\cr && \hspace{1cm}
   - {{47\,\pi \,\left( \gamma  + \ln 2 \right) }\over {35}} 
   - {{5\,i}\over {24}}\,{{\pi }^2}\,\left( \gamma  + \ln 2 \right)  - 
     {{47\,i}\over {35}}\,{{\left( \gamma  + \ln 2 \right) }^2} 
\cr && \hspace{1cm}
+    {{\pi \,{{\left( \gamma  + \ln 2 \right) }^2}}\over 4} + 
     {i\over 6}\,{{\left( \gamma  + \ln 2 \right) }^3} \Biggr) \Biggr],
\cr 
B^{in}_{3\omega} &=&
-{1\over \omega^2} e^{-i\epsilon(\ln 2\epsilon+\gamma)}
     \left(15-{3\over 4} i\epsilon\right)
\Biggl[1 + \epsilon \,\left( {{13\,i}\over 6} - {{\pi }\over 2} - 
     i\,\left( \gamma  + \ln 2 \right)  \right)  
\cr 
&& +
{{\epsilon }^2}\,\left( {{-26\,i}\over {21}}\,\pi  
+ {{5\,{{\pi }^2}}\over {24}} + 
     {{52\,\left( \gamma  + \ln 2 \right) }\over {21}} + 
     {i\over 2}\,\pi \,\left( \gamma  + \ln 2 \right)  - 
     {{{{\left( \gamma  + \ln 2 \right) }^2}}\over 2} \right)  
\cr 
&& + {{\epsilon }^3}\,\Biggl( {{3623\,i}\over {1080}} 
    + {{169\,\pi }\over {504}} + 
     {{481\,i}\over {1008}}\,{{\pi }^2} - {{{{\pi }^3}}\over {16}} + 
     {i\over 3}\,{\zeta(3)} + {{169\,i}\over {252}}\,
      \left( \gamma  + \ln 2 \right)  
\cr 
&& \hspace{1cm}
- {{39\,\pi \,\left( \gamma  + \ln 2 \right) }\over {28}} 
- {{5\,i}\over {24}}\,{{\pi }^2}\,\left( \gamma  + \ln 2 \right)  - 
     {{39\,i}\over {28}}\,{{\left( \gamma  + \ln 2 \right) }^2} 
\cr
& &\hspace{1cm}
+ {{\pi \,{{\left( \gamma  + \ln 2 \right) }^2}}\over 4} + 
     {i\over 6}\,{{\left( \gamma  + \ln 2 \right) }^3} \Biggr) \Biggr], 
\end{eqnarray}
where $\gamma$ and $\zeta(n)$ represent the Euler constant and 
the Riemann zeta function, respectively. 
In the expression of $B^{in}_{\ell\omega}$, we have assumed that $\omega$ is 
positive. The expression 
corresponding to negative $ \omega$ can be obtained by using 
the relation $B^{in}_{\ell-\omega}=B^{in*}_{\ell\omega}$.
For $\ell\ge 4$, we only need the forms of $B^{in}_{\ell\omega}$ 
valid up to $O(\epsilon)$, which can be found in Tagoshi and 
Sasaki. \cite{ref:TS} 

Using the above formulas, it is straightforward to obtain the luminosity of 
the gravitational waves. 
We define $\eta_{\ell,m}$ as 
\begin{equation}
{dE\over dt}\equiv {1\over 2}\left({dE\over dt}\right)_N 
\sum_{\ell,m} \eta_{\ell,m} , 
\end{equation}
where 
$(dE/dt)_N\equiv (32/5)(\mu/M)^2 v^{10}$ is the luminosity given by 
the well-known quadrupole formula. 
We show $\eta_{\ell,m}$ only for $m>0$, since $\eta_{\ell,m}=\eta_{\ell,-m}$ :

\begin{eqnarray}
\eta_{2,2}&=&1 - {{107\,{v^2}}\over {21}} + 4\,\pi \,{v^3} + 
  {{4784\,{v^4}}\over {1323}} - {{428\,\pi \,{v^5}}\over {21}} 
\nonumber\\
&+& {v^6}\,\left( {{99210071}\over {1091475}} - 
     {{1712\,\gamma }\over {105}} + {{16\,{{\pi }^2}}\over 3} - 
     {{3424\,\ln 2}\over {105}} - {{1712\,\ln v}\over {105}}
      \right)  
\nonumber\\
&+& {{19136\,\pi \,{v^7}}\over {1323}} 
\nonumber\\
&+& {v^8}\,\left( -{{27956920577}\over {81265275}} + 
     {{183184\,\gamma }\over {2205}} - 
     {{1712\,{{\pi }^2}}\over {63}} + 
     {{366368\,\ln 2}\over {2205}} + 
     {{183184\,\ln v}\over {2205}} \right)  
\nonumber\\
&+& {v^9}\,\left( {{396840284\,\pi }\over {1091475}} - 
     {{6848\,\gamma \,\pi }\over {105}} - 
     {{13696\,\pi \,\ln 2}\over {105}} - 
     {{6848\,\pi \,\ln v}\over {105}} \right)  
\nonumber\\
&+& {v^{10}}\,\left( {{187037845924}\over {6257426175}} - 
     {{8190208\,\gamma }\over {138915}} + 
     {{76544\,{{\pi }^2}}\over {3969}} 
\right.
\nonumber\\ & &\left. \hspace{1cm}
-    {{16380416\,\ln 2}\over {138915}} - 
     {{8190208\,\ln v}\over {138915}} \right)  
\nonumber\\
&+& {v^{11}}\,\left( {{-111827682308\,\pi }\over {81265275}} + 
     {{732736\,\gamma \,\pi }\over {2205}} 
\right.
\nonumber\\ & &\left. \hspace{1cm}
+ 
     {{1465472\,\pi \,\ln 2}\over {2205}} + 
     {{732736\,\pi \,\ln v}\over {2205}} \right), 
\\
\eta_{2,1}&=&
{{{v^2}}\over {36}} - {{17\,{v^4}}\over {504}} + 
  {{\pi \,{v^5}}\over {18}} - {{2215\,{v^6}}\over {254016}} - 
  {{17\,\pi \,{v^7}}\over {252}} 
\nonumber\\
&+& {v^8}\,\left( {{15707221}\over {26195400}} - 
     {{107\,\gamma }\over {945}} + {{{{\pi }^2}}\over {27}} - 
     {{107\,\ln 2}\over {945}} - {{107\,\ln v}\over {945}}
      \right)  
\nonumber\\
&-& {{2215\,\pi \,{v^9}}\over {127008}} 
\nonumber\\
&+& {v^{10}}\,
   \left( -{{6435768121}\over {57210753600}} + 
     {{1819\,\gamma }\over {13230}} - 
     {{17\,{{\pi }^2}}\over {378}} + 
     {{1819\,\ln 2}\over {13230}} + 
     {{1819\,\ln v}\over {13230}} \right)  
\nonumber\\
&+& {v^{11}}\,\left( {{15707221\,\pi }\over {13097700}} - 
     {{214\,\gamma \,\pi }\over {945}} - 
     {{214\,\pi \,\ln 2}\over {945}} - 
     {{214\,\pi \,\ln v}\over {945}} \right) , 
\\
\eta_{3,3}&=&{{1215\,{v^2}}\over {896}} - {{1215\,{v^4}}\over {112}} + 
  {{3645\,\pi \,{v^5}}\over {448}} + 
  {{243729\,{v^6}}\over {9856}} - 
  {{3645\,\pi \,{v^7}}\over {56}} 
\nonumber\\
&+& {v^8}\,\left( {{25037019729}\over {125565440}} - 
     {{47385\,\gamma }\over {1568}} + 
     {{3645\,{{\pi }^2}}\over {224}} - 
     {{47385\,\ln 2}\over {1568}} 
\right.
\nonumber\\
& &\left.
-    {{47385\,\ln 3}\over {1568}} - 
     {{47385\,\ln v}\over {1568}} \right)  
\nonumber\\
&+& {{731187\,\pi \,{v^9}}\over {4928}} 
\nonumber\\
&+& {v^{10}}\,\left( -{{2074855555161}\over {1381219840}} + 
     {{47385\,\gamma }\over {196}} - 
     {{3645\,{{\pi }^2}}\over {28}} + 
     {{47385\,\ln 2}\over {196}}
\right.
\nonumber\\
& &\left.
    +{{47385\,\ln 3}\over {196}} + 
     {{47385\,\ln v}\over {196}} \right)  
\nonumber\\
&+& {v^{11}}\,\left( {{75111059187\,\pi }\over {62782720}} - 
     {{142155\,\gamma \,\pi }\over {784}} - 
     {{142155\,\pi \,\ln 2}\over {784}} 
\right.
\nonumber\\
& &\left.
-    {{142155\,\pi \,\ln 3}\over {784}} - 
     {{142155\,\pi \,\ln v}\over {784}} \right) ,
\\
\eta_{3,2}&=&
{{5\,{v^4}}\over {63}} - {{193\,{v^6}}\over {567}} + 
  {{20\,\pi \,{v^7}}\over {63}} + 
  {{86111\,{v^8}}\over {280665}} - 
  {{772\,\pi \,{v^9}}\over {567}} 
\nonumber\\
&+& 
  {v^{10}}\,\left( {{960188809}\over {178783605}} - 
     {{1040\,\gamma }\over {1323}} + 
     {{80\,{{\pi }^2}}\over {189}} - 
     {{2080\,\ln 2}\over {1323}} - 
     {{1040\,\ln v}\over {1323}} \right) 
\nonumber\\
&+& 
  {{344444\,\pi \,{v^{11}}}\over {280665}} ,
\\
\eta_{3,1}&=&
{{{v^2}}\over {8064}} - {{{v^4}}\over {1512}} + 
  {{\pi \,{v^5}}\over {4032}} + {{437\,{v^6}}\over {266112}} - 
  {{\pi \,{v^7}}\over {756}} 
\nonumber\\
&+& {v^8}\,\left( -{{1137077}\over {50854003200}} - 
     {{13\,\gamma }\over {42336}} + {{{{\pi }^2}}\over {6048}} - 
     {{13\,\ln 2}\over {42336}} - {{13\,\ln v}\over {42336}}
      \right)  
\nonumber\\
&+& {{437\,\pi \,{v^9}}\over {133056}} 
\nonumber\\
&+& {v^{10}}\,
   \left( -{{38943317051}\over {5034546316800}} + 
     {{13\,\gamma }\over {7938}} - {{{{\pi }^2}}\over {1134}} + 
     {{13\,\ln 2}\over {7938}} + {{13\,\ln v}\over {7938}}
      \right)  
\nonumber\\
&+& {v^{11}}\,
   \left( {{-1137077\,\pi }\over {25427001600}} - 
     {{13\,\gamma \,\pi }\over {21168}} - 
     {{13\,\pi \,\ln 2}\over {21168}} - 
     {{13\,\pi \,\ln v}\over {21168}} \right) ,
\\
\eta_{4,4}&=&
{{1280\,{v^4}}\over {567}} - {{151808\,{v^6}}\over {6237}} + 
  {{10240\,\pi \,{v^7}}\over {567}} + 
  {{560069632\,{v^8}}\over {6243237}} - 
  {{1214464\,\pi \,{v^9}}\over {6237}} 
\nonumber\\
&+& {v^{10}}\,\left( {{36825600631808}\over {88497884475}} - 
     {{25739264\,\gamma }\over {392931}} + 
     {{81920\,{{\pi }^2}}\over {1701}} - 
     {{25739264\,\ln 2}\over {130977}} 
\right.
\nonumber\\
& &
\left.
 - {{25739264\,\ln v}\over {392931}} \right) 
+ {{4480557056\,\pi \,{v^{11}}}\over {6243237}} ,
\\
\eta_{4,3}&=&
{{729\,{v^6}}\over {4480}} - {{28431\,{v^8}}\over {24640}} + 
  {{2187\,\pi \,{v^9}}\over {2240}} + 
  {{620077923\,{v^{10}}}\over {246646400}} - 
  {{85293\,\pi \,{v^{11}}}\over {12320}} ,
\\
\eta_{4,2}&=&
{{5\,{v^4}}\over {3969}} - {{437\,{v^6}}\over {43659}} + 
  {{20\,\pi \,{v^7}}\over {3969}} + 
  {{7199152\,{v^8}}\over {218513295}} - 
  {{1748\,\pi \,{v^9}}\over {43659}} 
\nonumber\\
&+& {v^{10}}\,\left( {{9729776708}\over {619485191325}} - 
     {{25136\,\gamma }\over {2750517}} + 
     {{80\,{{\pi }^2}}\over {11907}} - 
     {{50272\,\ln 2}\over {2750517}} - 
     {{25136\,\ln v}\over {2750517}} \right) 
\nonumber\\
&+& {{28796608\,\pi \,{v^{11}}}\over {218513295}} ,
\\
\eta_{4,1}&=&
{{{v^6}}\over {282240}} - {{101\,{v^8}}\over {4656960}} + 
  {{\pi \,{v^9}}\over {141120}} + 
  {{7478267\,{v^{10}}}\over {139848508800}} - 
  {{101\,\pi \,{v^{11}}}\over {2328480}} ,
\\
\eta_{5,5}&=&
{{9765625\,{v^6}}\over {2433024}} - 
  {{2568359375\,{v^8}}\over {47443968}} + 
  {{48828125\,\pi \,{v^9}}\over {1216512}} 
\nonumber\\
& &+ 
  {{7060478515625\,{v^{10}}}\over {25904406528}} - 
  {{12841796875\,\pi \,{v^{11}}}\over {23721984}} , 
\\
\eta_{5,4}&=&
{{4096\,{v^8}}\over {13365}} - 
  {{18231296\,{v^{10}}}\over {6081075}} + 
  {{32768\,\pi \,{v^{11}}}\over {13365}} ,
\\
\eta_{5,3}&=&
{{2187\,{v^6}}\over {450560}} - 
  {{150903\,{v^8}}\over {2928640}} + 
  {{6561\,\pi \,{v^9}}\over {225280}} + 
  {{600654447\,{v^{10}}}\over {2665062400}} - 
  {{452709\,\pi \,{v^{11}}}\over {1464320}} ,
\\
\eta_{5,2}&=&
{{4\,{v^8}}\over {40095}} - {{15644\,{v^{10}}}\over {18243225}} + 
  {{16\,\pi \,{v^{11}}}\over {40095}} ,
\\
\eta_{5,1}&=&
{{{v^6}}\over {127733760}} - {{179\,{v^8}}\over {2490808320}} + 
  {{\pi \,{v^9}}\over {63866880}} 
\cr & &
+ 
  {{290803\,{v^{10}}}\over {971415244800}} - 
  {{179\,\pi \,{v^{11}}}\over {1245404160}} ,
\\
\eta_{6,6}&=&
{{26244\,{v^8}}\over {3575}} - 
  {{2965572\,{v^{10}}}\over {25025}} + 
  {{314928\,\pi \,{v^{11}}}\over {3575}} ,
\\
\eta_{6,5}&=&
{{244140625\,{v^{10}}}\over {435891456}}, 
\\
\eta_{6,4}&=&
{{131072\,{v^8}}\over {9555975}} - 
  {{4063232\,{v^{10}}}\over {22297275}} + 
  {{1048576\,\pi \,{v^{11}}}\over {9555975}}, 
\\
\eta_{6,3}&=&
{{59049\,{v^{10}}}\over {98658560}} ,
\\
\eta_{6,2}&=&
{{4\,{v^8}}\over {5733585}} - {{4\,{v^{10}}}\over {495495}} + 
  {{16\,\pi \,{v^{11}}}\over {5733585}} ,
\\
\eta_{6,1}&=&
{{{v^{10}}}\over {7192209024}} ,
\\
\eta_{7,7}&=&
{{96889010407\,{v^{10}}}\over {7116595200}} ,
\\
\eta_{7,5}&=&
{{6103515625\,{v^{10}}}\over {181330845696}} ,
\\
\eta_{7,3}&=&
{{1594323\,{v^{10}}}\over {205209804800}} ,
\\
\eta_{7,1}&=&
{{{v^{10}}}\over {5983917907968}} ,
\end{eqnarray}

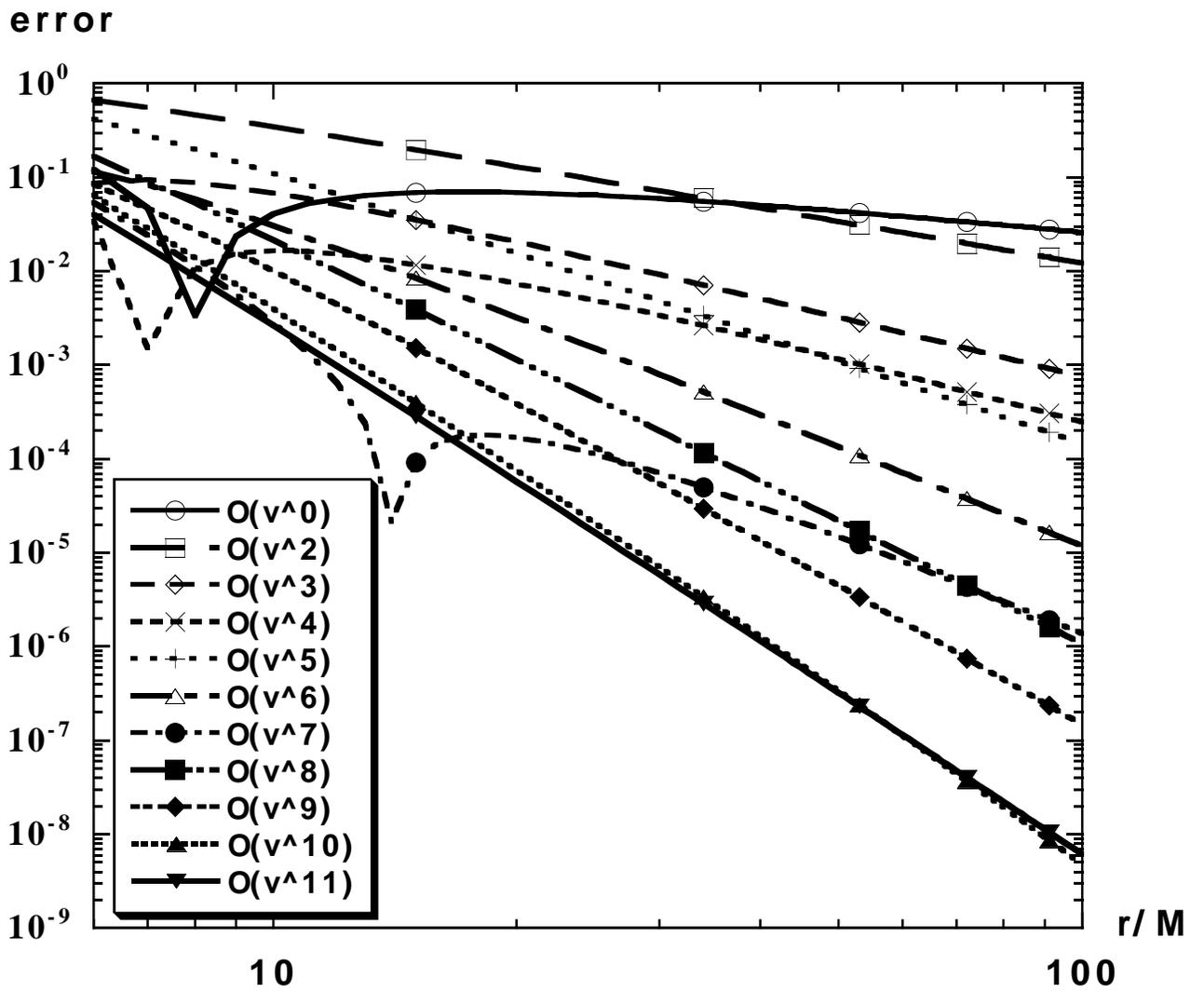
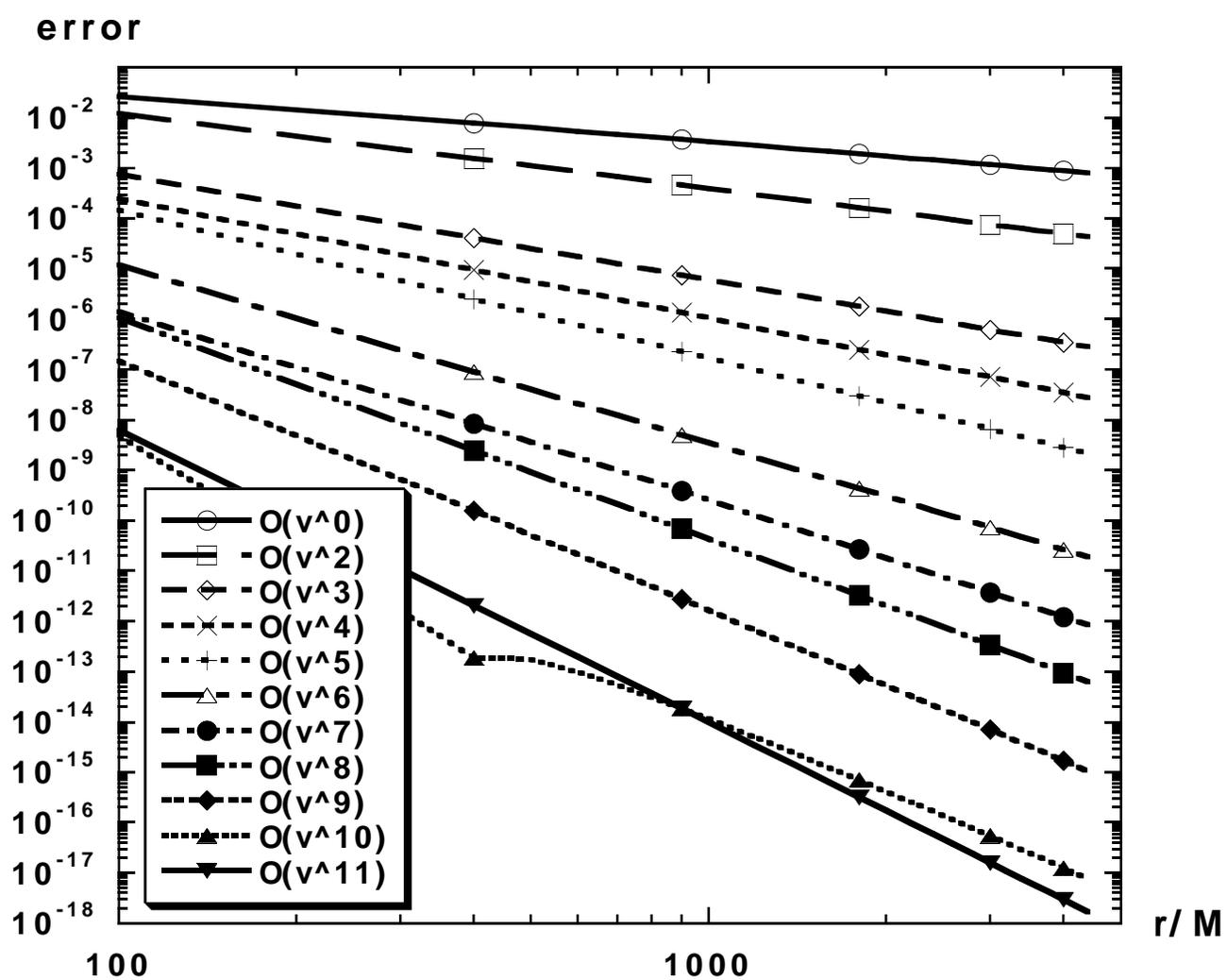
\end{document}